\documentstyle[12pt]{article}

\newcommand{\be}{\begin{equation}}
\newcommand{\ee}{\end{equation}}
\newcommand{\ba}{\begin{eqnarray}}
\newcommand{\ea}{\end{eqnarray}}
\begin{document}
\def\pct#1{\input epsf \centerline{ \epsfbox{#1.eps}}}

\begin{titlepage}
\hbox{\hskip 12cm ROM2F-96/12  \hfil}
\hbox{\hskip 12cm \today \hfil}
\vskip 1.5cm
\begin{center} 
{\Large  \bf  Completeness  Conditions
 for Boundary  Operators 
\vskip .6cm 
in  2D  Conformal  Field  Theory}
 
\vspace{1.8cm}
 
{\large \large G. Pradisi, \ \ A. Sagnotti \ \ and \ \ Ya.S. Stanev
\footnote{I.N.F.N.  Fellow,
on Leave from Institute for Nuclear Research and Nuclear Energy, Bulgarian 
Academy of Sciences, BG-1784 Sofia, BULGARIA.}}

\vspace{0.8cm}

{\sl Dipartimento di Fisica\\
Universit{\`a} di Roma \ ``Tor Vergata'' \\
I.N.F.N.\ - \ Sezione di Roma \ ``Tor Vergata'' \\
Via della Ricerca Scientifica, 1 \ \
00133 \ Roma \ \ ITALY}
\vspace{0.5cm}
\end{center}
\vskip 0.8cm
\abstract{In non-diagonal conformal models, the boundary fields are not 
directly related to the bulk spectrum.  We
illustrate some of their features by completing previous work of Lewellen on
sewing constraints for conformal theories in the presence of boundaries. 
As a result,
we include additional open sectors in the descendants of $D_{odd}$ $SU(2)$ WZW
models.  A new phenomenon emerges, the appearance of multiplicities and 
fixed-point
ambiguities in the boundary algebra not inherited from the closed
sector.  We conclude by deriving a set of polynomial equations,
similar to those satisfied by the fusion-rule coefficients $N_{ij}^k$,
for a new tensor $A_{a b}^i$ that determines the open spectrum.} 
 \vfill
\end{titlepage}
\addtolength{\baselineskip}{0.3\baselineskip} 

\vskip 24pt
\begin{flushleft}
{\large \bf Introduction}
\end{flushleft}

Cardy showed \cite{cardy} that in diagonal 
rational conformal models the admissible types of boundaries are in
one-to-one correspondence with the bulk fields. This observation has provided
a convenient setting \cite{bs,bps} for the proposal \cite{cargese}
of associating open descendants to closed oriented models, since different 
types of
boundaries correspond to different types of Chan-Paton groups \cite{cp}.  The
descendants \cite{pss1,pss2} of
$SU(2)$ WZW models \cite{wzw} revealed new aspects of the problem, most
notably the possibility of different Klein-bottle projections of the bulk 
spectrum,
fully determined by (a proper extension of) the crosscap constraint of ref.
\cite{fps}.

The descendants of non-diagonal models are harder to deal with,
since the simple correspondence between bulk and boundary sectors is lost in
this case\footnote{Rather unconventionally, by a
diagonal model we always mean one built with the charge-conjugation matrix $C$
of the fusion algebra.}.  In ref.
\cite{pss2} we solved the diophantine equations for the
non-diagonal $SU(2)$ WZW models, introducing $\rho + 3$ charge
sectors in the
$D_{odd}$ models with level $k=4 \rho + 2$.  This result was obtained
under the seemingly plausible assumption that all multiplicities in the annulus
amplitude be inherited from the closed spectrum, as for the $D_{even}$
models discussed in ref. \cite{pss1}.   The final
result, however, is rather puzzling, since the arguments of 
ref. \cite{bs} would
suggest a number of allowed boundary sectors equal to the 
number of characters paired with their conjugates by the
bulk $GSO$ projection, namely $2 \rho + 3$.

In this letter we reconsider the issue, by first completing and partly 
correcting
some previous interesting work of Lewellen \cite{lew} on sewing constraints for
conformal models in the presence of boundaries.  These constraints lead 
to a set of equations, sufficient to determine the vacuum channel of
the annulus amplitude, that are solved explicitly for the $D_{odd}$ models.  
Moreover, we prove that {\it the general solution to the
problem of classifying boundary conditions is given in terms of
an integer-valued tensor $A_{a b}^i$ satisfying a set of
polynomial equations}. These may be regarded 
as completeness conditions for the allowed boundaries.

For the boundary algebra, $A_{a b}^i$ plays a role similar
to that played by the fusion-rule coefficients $N_{ij}^k$ for the bulk algebra. 
The additional charges missed in ref. \cite{pss2} reveal a rather amusing and
unexpected feature: {\it the boundary algebra of non-diagonal models
can be extended even when the bulk algebra can not}, 
since the boundary states generically
correspond to (normalized) combinations of those allowed in the diagonal case. 
As a result, the annulus amplitude may contain some multiplicities that draw
their origin solely from the open sector.

 Aside from the application to open-string models, 
the construction of open descendants has some
interest for Statistical Mechanics \cite{lud}, as well as for the
emerging picture of non-perturbative string dynamics, where boundaries play an 
essential role \cite{polc,dbranes}, since the available choices 
of conformally invariant
boundary conditions determine the possible types of (generalized) $D$ branes.

\vskip 24pt
\begin{flushleft}
{\large \bf Sewing Constraints for the Annulus Amplitude}
\end{flushleft}

Sewing constraints for conformal models in the presence of boundaries were 
first discussed by Lewellen
\cite{lew}, but the original derivation contains some errors, 
and some of his final results need modifications. In the spirit of
ref. \cite{cargese}, let us
assume to  have solved the ``parent'' theory, so as to know its modular matrices
$T$ and $S$, the braid matrices $B$ and the duality matrices 
$F$ \cite{ms}, the fusion-rule coefficients 
$N_{ij}^k$ and the bulk OPE coefficients $C_{(i {\bar i})(j {\bar j})}^{(k {\bar
k})}$.  For a diagonal model, the open descendants are determined to a large 
extent by the construction of refs.
\cite{bs}, based on Cardy's ansatz \cite{cardy} for the annulus spectrum. 
Multiple Klein-bottle projections of the bulk spectrum were discussed in refs.
\cite{pss1,pss2}, where (a proper extension of) the ``crosscap''
constraint of ref.
\cite{fps} was used to determine them for all $SU(2)$ WZW
models\footnote{In these models, the open spectrum contains a simple current
that connects pairs of different charge assignments.}. 
For non-diagonal models, one does
not have so far an equivalent recipe, and the main purpose 
of this letter is to set the stage for this more general case.

Denoting the ``bulk fields'' of the theory by
$\phi_{i,{\bar i}}$ and the ``boundary fields'' by ${\psi_i}^{ab}$, one has the
usual bulk OPE, as well as the boundary OPE
\be
{\psi_i}^{ab} \ {\psi_j}^{bc} \sim \sum_l \ C_{ijl}^{abc} \ \psi_l^{ac} \qquad .
\label{bope}
\ee
Additional data of the descendant models are the normalizations of two-point
functions, ${\alpha_i}^{ab}$, defined by
\be
< {\psi_i}^{ab} ( x_1 ) \ {\psi_i}^{ba} ( x_2 ) > \ = \ {{{\alpha_i}^{ab}} \over
{(x_{12})^{2
\Delta_i}}} \quad ,
\label{norm}
\ee
where for $SU(2)$ WZW models
\be
{\alpha_i}^{ab}  \ = \ {\alpha_i}^{ba} \ (-1)^{2 I_i} \quad ,
\label{alphas}
\ee
with $I_i$ the isospin of $\psi_i$. While restricting our attention to 
$SU(2)$ WZW
models, we would like to point out that all our formulas may be turned into
corresponding ones for minimal models, provided all isospin-dependent 
factors are set
to one.  Making use of eq. (\ref{bope}) in the three-point functions
of boundary operators
$<{\psi_i}^{ab} {\psi_j}^{bc} {\psi_l}^{ca}>$ and $<~{\psi_j}^{bc}
{\psi_l}^{ca} {\psi_i}^{ab}>$ leads to
\be
C_{ijl}^{abc} \ {\alpha_l}^{ac} \ = \ C_{jli}^{bca} \ {\alpha_i}^{ab} \qquad
{\rm and} \qquad
C_{jli}^{bca} \ {\alpha_i}^{ba} \ = \ C_{lij}^{cab} \ {\alpha_j}^{bc} \quad .
\ee
These two relations may be connected using eq. (\ref{alphas}), and one finally
obtains
\be
C_{ijl}^{abc} \ {\alpha_l}^{ac} \ = \  (-1)^{2 I_i}  \
C_{lij}^{cab} \ {\alpha_j}^{bc}	\qquad .
\label{constralpha}
\ee
Moreover, the proper behavior of the identity requires that
\be
C_{i{\bf 1}i}^{abb} \ = \ 1	\qquad 	{\rm and} \qquad < {\bf 1}^{aa} > \ = \
\alpha_{\bf 1}^{aa}  \qquad , \label{identity}
\ee
while all other one-point functions of boundary fields
vanish.

One may then proceed to consider amplitudes 
$<~\psi_i^{ab} \ \psi_j^{bc} \ \psi_k^{cd} \ \psi_l^{da} >$ 
for four boundary operators.
Demanding that their  $s$ and $u$-channel expansions coincide 
(``planar duality'' for open four-point amplitudes) yields
\be
\sum_p \ C_{ijp}^{abc} \ C_{klp}^{cda} \ \alpha_p^{ac} \ S_p(i,j,k,l) \ = \
\sum_q \ C_{jkq}^{bcd} \ C_{liq}^{dab} \ \alpha_q^{bd} \ U_q(i,j,k,l) \quad ,
\label{opplanar}
\ee
and relating the $u$-channel blocks to the $s$-channel
ones by the fusion matrix $F$
\be
U_q \ = \ \sum_p \ F_{qp} \ S_p	\qquad 
\label{duality}
\ee
turns eq. (\ref{opplanar}) into
\be
C_{ijp}^{abc} \ C_{klp}^{cda} \ \alpha_p^{ac} \ = \
\sum_q \  C_{jkq}^{bcd} \ C_{liq}^{dab} \ \alpha_q^{bd} \ 
F_{qp} (i,j,k,l) \quad ,
\label{constr4o}
\ee
a quadratic constraint for the boundary OPE
coefficients
$C_{ijk}^{abc}$ and the normalizations $\alpha_i^{ab}$  of 
the two-point functions of boundary fields.

The last crucial ingredient of the construction, 
introduced in refs. \cite{cardylew,lew},
is the OPE for bulk fields in front of a boundary.  
This corresponds to a familiar
intuitive picture: when a bulk field approaches a boundary, 
the result should be expressible solely in terms of boundary fields.  Thus,
\be
\phi_{i, {\bar i}} \ \sim \ \sum_j \ C_{(i,{\bar i})j}^a \
{\psi_j}^{aa} \quad ,
\label{bulktobound}
\ee
where the proper behavior of the identity requires that
\be
 C_{({\bf 1},{\bf 1}){\bf 1}}^a \ = \ 1	\qquad .
\ee

One may then proceed to consider amplitudes for one bulk field and two
boundary fields.  There are two ways of computing
$< \phi_{(i,{\bar i})} \ \psi_j^{ba} \ \psi_k^{ab} >$, according to which 
portion of the boundary the bulk field faces, and the resulting condition is
\be
\sum_l \
C_{(i,{\bar i}) l}^b \ C_{ljk}^{bba} \ \alpha_k^{ba} \ S_l(i,{\bar i},j,k)
\ = \ 
\sum_n \ C_{(i,{\bar i})n}^a \ C_{jnk}^{baa} \ \alpha_k^{ba} 
\ U_n(j,i,{\bar i},k)
\quad .
\label{opclosed1}
\ee
\vskip 12pt
\input epsf \centerline{ \epsfbox{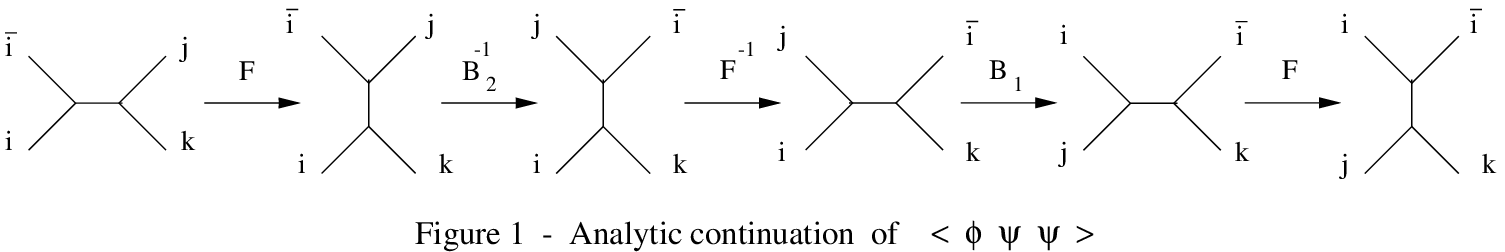}}
\vskip 12pt
\noindent
Finding the proper relation between the blocks $U_n$ and $S_l$
involves some delicate analytic continuations, and we disagree with the 
final result of ref. \cite{lew}.  In order to derive this sewing
constraint, we shall reduce the relevant transformation to
a sequence of elementary
moves, as in fig. 1.  These comprise two basic operations, the
braidings $B_i$ of pairs of nearby operators and the fusion $F$ \cite{ms},
and the result reads
\ba
U_n (j,i,{\bar i},k) \ &=& \ \sum_{m,r,s,p,l} \ F_{nm}(j,i,{\bar i},k) \ 
( B_1 )_{mr} (i,j,{\bar i},k) \ F^{-1}_{rs}(i,j,{\bar i},k) \nonumber \\ 
& &( B_2 )^{-1}_{sp} ( i,{\bar i},j,k ) \ F_{pl} (i,{\bar i},j,k) \
S_l (i,{\bar i},j,k) \qquad .
\ea
Substituting in eq. (\ref{opclosed1}) and recalling that in this case the
braid matrices $B_i$ are diagonal \cite{rst} yields
the final form of the constraint,
\ba
C_{(i,{\bar i})l}^b \ C_{j k l}^{bab} \ {\alpha_l}^{bb} \ &=& \
\sum_{m,n,p} \ C_{(i,{\bar i})n}^a \ C_{kjn}^{aba} \ {\alpha_n}^{aa}
(-1)^{(I_i - I_{\bar i} + 2 I_j + I_p - I_m )} \
e^{- i \pi ( \Delta_i - \Delta_{\bar i} - \Delta_m + \Delta_p )} \nonumber \\ 
& &{F_{nm}}(j,i,{\bar i},k) \ {F^{-1}}_{mp}(i,j,{\bar i},k) \
F_{pl}(i,{\bar i},j,k)
\qquad .
\label{opclosed2}
\ea

Similar considerations apply to the last constraint of ref. \cite{lew}.  This
results from the comparison between two different definitions of three-point
amplitudes for two bulk fields and one boundary field, $< \phi_{(i,{\bar i})} \
\phi_{(j,{\bar j})} \ {\psi_k}^{aa} >$. These are effectively chiral five-point
amplitudes, and are thus more complicated than the previous ones. 
In this case, the
first definition uses the bulk OPE,  while the second definition uses a pair
of bulk-boundary OPE's. 
Demanding that the two resulting expressions coincide gives
\be
\sum_{p,q} \ C_{(j,{\bar i})(i,{\bar j})}^{ (p,{\bar q})} \
C^a_{(p,{\bar q})k} \ \alpha_k^{aa} \ Y_{p{\bar q}}(j,i,{\bar i},{\bar j},k)
\ = \ \sum_{p,q} \ C^a_{(i,{\bar i})p} \ C^a_{(j,{\bar j})q} \ 
C_{pqk}^{aaa} \ \alpha_k^{aa} \ X_{pq}(i,{\bar i},j,{\bar j},k)
\quad .
\label{secondform}
\ee
\vskip 12pt
\input epsf \centerline{ \epsfbox{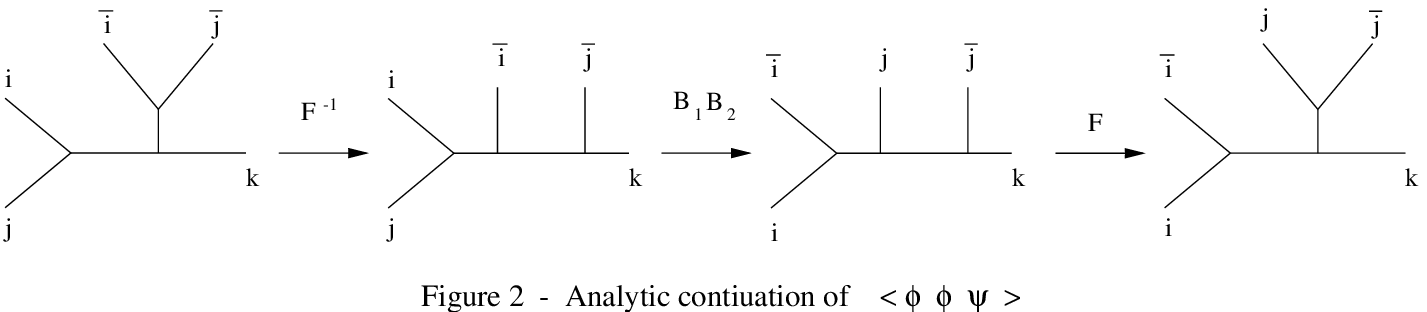}}
\vskip 12pt
\noindent
Again, relating the two expressions requires a careful analytic
continuation, that may be reduced to the sequence of elementary moves
displayed in fig. 2. The resulting constraint, present only if 
$\alpha_k^{aa}$ does not vanish, reads
\ba
C_{(j,{\bar i})(i,{\bar j})}^{ (p,{\bar q})} \ C^a_{(p,{\bar q})k} \ &=& \
\sum_{r,s,t} \ (-1)^{( I_j - I_t + I_r )} \ 
e^{- i \pi ( \Delta_j - \Delta_t + \Delta_r )} \
C^a_{(i,{\bar i}) r} \ C^a_{(j,{\bar j})s} \ C_{rsk}^{aaa} \nonumber \\
& &F_{st}(r,j,{\bar j},k) \ F_{rp}(j,i,{\bar i},t) \ 
F_{t{\bar q}}^{-1}(p,{\bar i},{\bar
j},k)  \quad . \label{opclosed3}
\ea
For $SU(2)$ WZW models the bulk OPE coefficients satisfy
\be
C_{(j,{\bar i})(i,{\bar j})}^{ (p,{\bar q})} \ = \ (-1)^{I_i + I_j - I_p} \
C_{(i,{\bar i})(j,{\bar j})}^{ (p,{\bar q})} \qquad ,
\ee
and are normalized in a different fashion with respect to ref. \cite{pss2},
so that now
\be
C_{(i,k)(j,l)}^{ (p,q)} \ = \ \epsilon_{(i,j)(k,l)}^{ (p,q)} 
\sqrt{{{C_{ijp} C_{klq}} \over {C_{pp{\bf 1}} C_{qq{\bf 1}}}}} \quad ,
\ee
since this choice simplifies our final expressions.  The rather complicated
expressions for $F$ and $C_{ijk}$ depend on
normalization choices.  We follow ref. \cite{rst}, while some of the
relevant formulas may also be found in ref. \cite{pss2}. 
Moreover, the $\epsilon$'s are
signs present only for the non-diagonal models. In the next section we shall
specify them explicitly for the cases of interest.

We have thus completed the derivation of the sewing constraints.  
The resulting relations (and
their solutions) differ from those obtained in ref. \cite{lew} 
even for the simplest
case of the Ising model. In the $SU(2)$ WZW models a subset 
of these constraints, sufficient
to determine the annulus vacuum amplitude, decouples.  In order 
to elucidate this point, let us confine our attention to the amplitude
$< \phi_{(i,{\bar i})} \ \phi_{(j,{\bar j})} \ {\bf 1}^{aa} >$, whereby eq.
(\ref{opclosed3}) becomes
\ba
C_{(j,{\bar i})(i,{\bar j})}^{ (q,{\bar q})} \ C^a_{(q,{\bar q}){\bf 1}} 
\ \alpha_{\bf 1}^{aa} \ &=& \
\sum_{p} \ (-1)^{( I_j - I_{\bar j} + I_p )} \ 
e^{- i \pi ( \Delta_j - \Delta_{\bar j} + \Delta_p )} \nonumber \\
& &\alpha_p^{aa} \
C^a_{(i,{\bar i})p} \ C^a_{(j,{\bar j})p} \
F_{pq}(j,i,{\bar i},{\bar j})   \quad . \label{opclosed4}
\ea
Multiplying by $F^{-1}_{qr}(j,i,{\bar i},{\bar j})$, singling out 
the identity ({\it i.e.} choosing $r={\bf 1}$), defining
\be
{B}_{(i,{\bar i})}^a = C^a_{(i,{\bar i}){\bf 1}}
\label{vacuum}
\ee
and using the explicit expressions for fusion matrices and structure 
constants of ref. \cite{rst} yields a set of relations 
involving {\it only} the $B$'s, 
the basic ingredients of the annulus vacuum channel.  
Since the $\alpha_{\bf 1}^{aa}$ never vanish, one obtains the simple constraint 
\be
{B}_i^a \ {B}_j^a \ = \ \sum_l \ 
\epsilon_{ij}^l \ N_{ij}^l \ {B}_l^a
\qquad ,
\label{neat}
\ee
where we have expressed the restriction of the
sum to all terms in the ``fusion range'' of $i$ and $j$ via the fusion-rule
coefficients $N_{ij}^k$ and we have simplified the notation, replacing
every pair of (coincident) indices with a single one.
  
Once the solution to eq. (\ref{neat}) has been found, the
vacuum-channel annulus amplitude is
\be
{\tilde A} \ = \ {1 \over 2} \sum_i 
{{\chi_i} \over {[2 I + 1]}} \ {\left( {\sum_a
{B}_i^a \ n^a \ \alpha_{\bf 1}^{aa}} \right)}^2 \quad , \label{atilde}
\ee
since we have normalized the two-point
functions of the bulk fields to their quantum dimensions $[ 2 I + 1 ]$
while, in general, the direct channel annulus amplitude is of the form
\be
A \ = {1 \over 2} \ \sum_{a b i} A_{a b}^i \ n^{a} \
n^{b} \ \chi_i	\qquad .
\label{annulusgen}
\ee
In open-string theories, the non-negative integers $A_{a b}^i$ encode the
properties of the open sector.  In rational conformal models, 
they determine the set
of conformally invariant boundary conditions or, equivalently, they
count the boundary fields $\psi_i^{a b}$.

Although we have derived eq. (\ref{neat}) in the context of $SU(2)$ 
WZW models, a generalization holds in all cases.  For instance, 
in toroidal models one can show that the most general
modification of the vacuum-channel coefficients involves 
multiplicative phases.  These are just the Wilson lines of ref.
\cite{bps} that implement in open-string models the construction of ref.
\cite{nsw}.
\vskip 24pt 
\begin{flushleft}
{\large \bf Application to the $D_{odd}$ $SU(2)$ WZW Models}
\end{flushleft}

It is instructive to apply the results of the preceding section to the 
$D_5$ model, the simplest
non-diagonal $SU(2)$ WZW model. This will allow us to 
reconsider the construction
of ref. \cite{pss2}, that actually turns out to be incomplete.  In 
that paper we found by brute force only four boundary
sectors, a somewhat surprising result in
view of the conventional wisdom, that leads one to expect five sectors, as
many as the
bulk fields allowed in the annulus vacuum channel.  
Indeed, the ${B}_{i}^a$ determine
the number of independent combinations of the $n$'s, 
and thus the independent charge
sectors of the model, but they also determine the one-point 
functions of the bulk fields in
front of a boundary, via their products with the
$\alpha$'s. Since these one-point functions are essentially chiral two-point
functions on the sphere, non-vanishing
results obtain {\it only} for the fields that the
$GSO$ projection of the bulk spectrum mixes with their charge conjugates. 

The spectrum of the $D_5$ model in the $ADE$ classification \cite{ciz}
is described by
\be
T \ = \ |\chi_1|^2 \ + \ |\chi_3|^2 \ + \ |\chi_5|^2 \ + \ |\chi_7|^2 \ + \
\ |\chi_4|^2 \ + \chi_2 {\bar \chi}_6 + \chi_6 {\bar \chi}_2 \quad
, \label{tord3}
\ee
where the subscript of $\chi_i$ is related to its isospin $I$ by
$i = 2I+1$. Since all these characters are self-conjugate, 
one would indeed expect five charge sectors, whereas 
in ref. \cite{pss2} we could only identify four of them.  

One may apply eq. (\ref{neat}) to this case,
noting that the only difference between the two systems of quadratic
equations  for the $A_6$ and $D_5$ models lies in the signs
$\epsilon_{ij}^l$, all equal to one in the diagonal model,
and the two subsystems for the integer-isospin coefficients
are identical.  For the diagonal $A_6$ model, the system has a total of seven
distinct solutions.
Each complete choice of coefficients determines, according 
to eq. (\ref{atilde}),
the contribution of one type of Chan-Paton charge to the vacuum amplitude. 
Strictly speaking, this would also require the $\alpha$'s, that in the diagonal
model can be determined as in ref. \cite{cardy}. More generally, one may 
determine them completely turning the
annulus amplitude to the direct channel by a modular $S$ 
transformation and requiring that the coefficients be (half)integer.  
For the diagonal
model one thus recovers the seven types of charges of ref. \cite{pss1}.  
The resulting correspondence
between charge types and open-string sectors is displayed in the first column of
the table.
\vskip 12pt
\noindent
\begin{tabular}{|r||r|r|r|r||r|r|r||r|}
\hline
\multicolumn{9}{|c|}{$B$ coefficients for the $A_6$ and $D_5$ models}
\\ \hline\hline
{Op's} & ${{B}_1}$ & ${{B}_3}$ & ${{B}_5}$& ${{B}_7}$ &
${{B}_4^{(A_6)}}$ & ${{B}_2}$ & ${{B}_6}$ & ${{B}_4^{(D_5)}}$
\\ \hline\hline
$({1 \over 2},{5 \over 2})$ & $1$ & $1$ & $-1$ & $-1$ & $0$ 
& $\pm \sqrt{2}$ & $\mp \sqrt{2}$ & $0$
\\ \hline
$({3 \over 2})$ & $1$ & $-1$ & $1$ & $-1$ & $0$ & $0$ & $0$ & $\pm 2$
\\ \hline
$(0,3)$ & $1$ & ${1+\sqrt{2}}$ & ${1+\sqrt{2}}$ & $1$ & 
$\pm \sqrt{2({2+\sqrt{2}})}$ & 
$\pm \sqrt{2+\sqrt{2}}$ & $\pm \sqrt{2+\sqrt{2}}$ & $0$
\\ \hline
$(1,2)$ & $1$ & ${1-\sqrt{2}}$ & ${1-\sqrt{2}}$ & $1$ & 
$\mp \sqrt{2({2+\sqrt{2}})}$ & $\pm \sqrt{2-\sqrt{2}}$ & 
$\pm \sqrt{2 -\sqrt{2}}$ & $0$
\\ \hline
\end{tabular}
\vskip 12pt
Extending this analysis to the $D_5$ model is particularly rewarding.  In this
case both ${B}_{2}$ and ${B}_{6}$ vanish for the
general argument discussed above, while the new equations for
${B}_{4}$,
\ba
& &{B}_{4} \ {B}_{2I+1} \ = \ (-1)^I {B}_{4} \quad ,\nonumber \\
& &{B}_{4} \ {B}_{4} \ = \ {B}_{1} - {B}_{3} + 
{B}_{5} - {B}_{7} \quad ,
\label{offidiagb}
\ea
involve some additional signs introduced by the $\epsilon$'s. 
${B}_{4}$ thus
vanishes, unless ${B}_{2I+1}$ is precisely
$(-1)^I$.  This occurs in the third row of the table, but in this case
there are two solutions,
${B}_{4} = \pm 2$, as displayed in the last column. 
A closer inspection of the table reveals the correspondence between the
boundary states of the two models, determined by the condition that 
both ${B}_{2}$ and
${B}_{6}$ vanish in the non-diagonal case.  Some of the new boundary states are
created from the vacuum by the following linear combinations of the
boundary operators of the diagonal model:
\ba
\xi_1 \ &=& \ {1 \over {\sqrt{2}}} ( \psi_{2}^{2,1} + \psi_{6}^{6,1}) \nonumber
\\
\xi_2 \ &=& \ {1 \over {\sqrt{2}}} ( \psi_{1}^{1,1} + \psi_{7}^{7,1}) \nonumber
\\
\xi_5 \ &=& \ {1 \over {\sqrt{2}}} ( \psi_{3}^{3,1} + \psi_{5}^{5,1}) \quad 
\label{corresp}
\ea
where, again, all labels on the {\it r.h.s.} correspond to $2 I + 1$.
The correspondence, however, is only partial, since
there are now {\it two} boundary 
sectors $\xi_{3}$ and $\xi_{4}$ corresponding to the middle field
$\psi_{4}$~! A fixed-point ambiguity, not present in the spectrum 
of bulk fields, has emerged in the set of
boundary fields.  This is not the
only surprise, since we did get these charges in ref. \cite{pss2}.  What we 
missed there
was the charge sector corresponding to $\xi_5$, since we allowed
no multiplicities in the boundary fusion
algebra, and thus in the direct-channel annulus amplitude.  Multiplicities are
indeed present, as may be foreseen
from eq. (\ref{corresp}), since the fusion of
$\xi_5$ is 
\be
[ \xi_5 ] \ \times \ [ \xi_5 ] \ = \ [ \xi_2 ] \ + \ 2 \ [ \xi_5 ] \quad .
\label{bondaryfuse}
\ee 
Thus, as compared to the diagonal case, the algebra of boundary 
operators is an extended
algebra.  This is rather amusing, since the simple current in this case 
has dimension
3/2, and therefore does not extend the bulk algebra. As a result, 
the complete open sector 
of the model with ``real'' charges is described by
\ba
\lefteqn{A \ = \ {1 \over 2} \ \biggl( 
\chi_1 ( l_1^2 + l_2^2 + l_3^2 + l_4^2 + l_5^2 ) \ + \
(\chi_2 + \chi_6 )( 2 l_1 l_2  + 2 l_1 l_5 + 2 l_3 l_5 + 2 l_4 l_5) \ +}
\nonumber
\\ & & \chi_3 ( l_1^2 + 2 l_1 l_3 + 2 l_1 l_4 + 2 l_3 l_4 + 2 l_2 l_5 + 2 l_5^2)
 \ + \nonumber \\
& & \chi_4 ( 4 l_1 l_5 + 2 l_2 l_3 + 2 l_3 l_5 + 2 l_2 l_4 + 2 l_4 l_5 )  \ +
\label{adr} \\  
& &\chi_5 (l_1^2  + l_3^2  + l_4^2 + 2 l_5^2 + 2 l_1 l_3 + 2 l_1 l_4 + 2 l_2
l_5 ) \ +
\
\chi_7 ( l_1^2 + l_2^2 + l_5^2 + 2 l_3 l_4 ) \biggr) \nonumber \quad , 
\ea
and 
\ba
M &=& \pm \ {1 \over 2} \ \biggl( {\hat \chi}_1 ( l_1 - l_2 + l_3 +
l_4 - l_5 ) \ + \ {\hat \chi}_3 ( - l_1 + 2 l_5 ) \ +
\nonumber \\
& & \qquad {\hat \chi}_5 (l_1 + l_3 + l_4 ) \ + 
\ {\hat \chi}_7 ( l_1 + l_2 + l_5 ) \biggr)  \label{mdr} 
\ea
where the labels of the charges correspond to those of the $\xi$'s.
Indeed, the new charge $l_5$ has multiplicities both in
the annulus and in the M\"obius amplitude. The model with complex charges
involves similar modifications.

These results may be extended to all $D_{odd}$ models,
thus allowing for a total of $2 \rho + 3$ charges. The
resulting assignments may be described rather neatly in terms of an auxiliary
diagonal model so that, in the notation of eq. (37) in ref. \cite{pss2}, 
the complete embedding for the case of real charges is
\be
n_a \ = \ n^{+}_a \ + \ n^{-}_a \ + \ {i \over {2 \sqrt{\rho +1}}} \  
O_a (-1)^{a-1 \over 2} \
(l_{\rho +2} - l_{\rho +3} ) \qquad  ({\rm a \ = \ } 1 , \ ... \ , k+2) 
\quad , \label{reduction1}
\ee
where $O_a$ denotes the projector on odd $a$ and the $n^{\pm}$ satisfy the 
relations
\ba
& &n^{\pm}_{{k+2 \over 2}+b} \ = \ n^{\pm}_{{k+2 \over 2}-b} \quad
, \qquad
n^{\pm}_{{k+2 \over 4}+b} \ = \ \pm \ n^{\pm}_{{k+2 \over 4}-b} 
\qquad ( b \ge 1 ) \quad , \nonumber \\
& &n^{-}_b \ = \ - \ {l_{b+1} \over 2} \quad ,  
\qquad n^{+}_b \ = \ {l_{\rho+3+b} \over
2} \quad , \qquad ( 1 \le b \le \rho ) \quad ,  \label{reduction2} \\
& &n^{-}_{k+2 \over 2} \ + \ n^{+}_{k+2 \over 2} \ = \ l_1 \quad , \qquad 
n^{-}_{k+2 \over 4} \ + \ n^{+}_{k+2 \over 4} \ = \ {1 \over 2} 
\left( l_{\rho +2} + l_{\rho +3} \right) \quad .
\nonumber
\ea
\vskip 24pt
\begin{flushleft}
{\large \bf Completeness Conditions for Boundary Operators}
\end{flushleft}

A closer inspection of eq. (\ref{adr}) reveals a very interesting property. 
Namely, it may be verified that the 
non-negative integers $A_{a b}^i$ defined in eq. (\ref{annulusgen}) 
satisfy two sets of polynomial equations involving also the fusion-rule 
coefficients $N_{ij}^k$,
\ba
\sum_{b} \ {A^i}_{a}^{b} \ A_{b c}^j \ &=& \
\sum_k N^{ij}_k \  A_{a c}^k \qquad  , \label{conj1} \\	
\sum_{i} \ A_{i a b} \ A_{c d}^i \ &=& \
\sum_{i} \ A_{i a c} \ A_{b d}^i \qquad , \label{conj2}
\ea
while omitting the $l_5$ terms would violate eq. (\ref{conj1}).
Upper and lower boundary indices are to be 
distinguished whenever complex charges 
(corresponding to oriented boundaries) are present. The matrix
$(A_1)_{a b} = (A_1)^{a b}$ is a metric for the boundary indices, since
it follows from eq. (\ref{conj1}) that  
$\sum_b A_{i a b} \ {A_1}^{b c} = {A_i}_a^c$,
while $(A_1)_a^b = \delta_a^b $.
In diagonal models, where $A$ coincides with $N$, these equations reduce to 
the Verlinde algebra. 

One can prove that these polynomial equations hold for all 
rational conformal field
theories if the boundary states $| b >$ form a complete set.  
To this end let us recall that, by definition $A_i^{a b}$ 
counts the number of boundary operators ${\psi}_i^{a
b}$.  These, however, are determined by a boundary algebra 
(Virasoro, current, or some other
extended algebra) that has the same central charges, 
and hence the same representations,
as the bulk one. Therefore,  $A_i^{a b}$ also counts the number 
of different couplings 
$< a | \phi_i | b >$, where $| a >$ and $| b >$ denote boundary states.  
We have labeled
the two-dimensional fields $\phi_{i ,{ \bar i}}$ by their chiral weights that,
once all fixed-point ambiguities are resolved, determine the antichiral 
ones.  Eq.
(\ref{conj1}) then follows if one computes the number of
couplings $< a | \phi_i \ \phi_j | b >$ in two ways, by using
the bulk fusion rules or by expanding in terms of a complete set of
boundary states:
\ba
< a | \phi_i \ \phi_j | c > \ &=& \
\sum_l \ N^{ij}_l \ < a | \phi_l | c > \nonumber \\ &=& \
\sum_{b} \ < a | \phi_i | b > < b | \phi_j | c > \ = \ \sum_{b} \ {A^i}_{a}^{b}
\ A_{b c}^j \quad .
\label{proof}
\ea
Finally, eq. (\ref{conj2}) follows from the general structure 
of the vacuum-channel annulus
amplitude of eq. (\ref{atilde}), which implies that the bilinears are
totally symmetric in their boundary indices.

Eqs. (\ref{conj1}) and (\ref{conj2}) do not determine completely the matrices
$A_i^{a b}$, since they contain only chiral data.  Another crucial ingredient of
the construction, the torus modular invariant,
determines the non-vanishing disk one-point functions, and thus the range of the
boundary indices, generally smaller than the range of the bulk indices.
We have verified that, with this proviso, eqs. (\ref{conj1}) 
and (\ref{conj2}) have a unique solution, up to a 
relabeling of the boundary indices, for all the models
that we have analyzed, and in particular for minimal and $SU(2)$ WZW models.
For the $D_{odd}$ series, the solution is implied by eqs. (\ref{reduction2}).
In general, one can pick a subset of linearly independent $A$ matrices, but 
when the boundary algebra is extended (as in the $D_{odd}$
models) one can not interpret them as the fusion-rule coefficients
of any conformal model.
Some useful corollaries can be obtained even without solving 
eqs. (\ref{conj1}) and (\ref{conj2}) explicitly. 
For instance, for abelian fusion
rules each sum on the right hand side of eq. (\ref{conj1}) reduces to 
only one term, and
this implies that no multiplicities larger than one are present in this case.

Matters look deceptively simpler if one introduces a graphical notation 
for $A_{a b}^i$ and
$N_{ij}^k$, where boundary indices correspond to dashed lines and
bulk indices
correspond to continuous lines (fig. 3).  Then eqs. (\ref{conj1}) 
and (\ref{conj2}),
together with the Verlinde algebra, express the ``planar duality'' 
of all four-point
amplitudes built out of these two kinds of three-point vertices.

\vskip 12pt
\input epsf \centerline{ \epsfbox{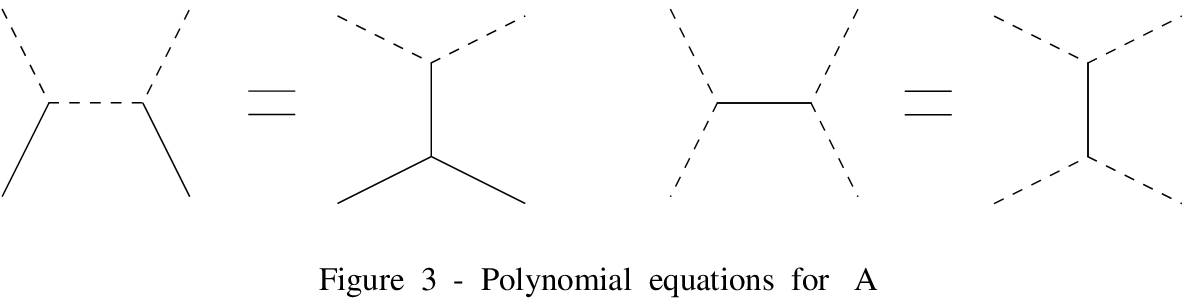}}
\vskip 12pt

Since the available choices of conformally invariant
boundary conditions determine the possible types of (generalized) $D$ branes,
the completeness conditions are expected to play a role in the emerging
picture of non-perturbative string dynamics \cite{polc,dbranes}. 

A.S. would like to thank K. Dienes, E. Witten, S. Mathur, B. Zwiebach,
D. Anselmi and C. Vafa for their kind hospitality at the Institute for
Advanced Study, Princeton, M.I.T. and Harvard University, and
Ya.S.S. would like to acknowledge the kind hospitality of
the Erwin Schr\"odinger Institute in Vienna while this work was 
being completed.
This work was supported in part by E.E.C. Grant CHRX-CT93-0340.


\begin{thebibliography}{99}
\bibitem{cardy} J.L. Cardy, {\sl Nucl. Phys.} {\bf B324} (1989) 581.
\bibitem{bs} M. Bianchi and A. Sagnotti, {\sl Phys. Lett.} {\bf B247}
(1990) 517, \\ {\sl Nucl. Phys.} {\bf B361} (1991) 519.
\bibitem{bps} M. Bianchi, G. Pradisi and A. Sagnotti, {\sl Phys. Lett.}
{\bf B273} (1991) 389,\\ {\sl Nucl. Phys.}  {\bf B376} (1992) 365.
\bibitem{cargese} A. Sagnotti, {\it in} ``Non-Perturbative Quantum 
Field Theory'',\\
eds. G. Mack et al (Pergamon Press, 1988), p. 521.
\bibitem{cp} J.E. Paton and H.M. Chan,  {\sl Nucl. Phys.} 
{\bf B10} (1969) 516;\\
J.H. Schwarz, in ``Current Problems in Particle Theory'',\\ 
Proc. J. Hopkins Conf. {\bf 6} (Florence, 1982);\\
N. Marcus and A. Sagnotti, {\sl Phys. Lett.} {\bf B119} (1982) 97, {\bf B188}
(1987) 58.
\bibitem{pss1} G. Pradisi, A. Sagnotti and Ya.S. Stanev, 
{\sl Phys. Lett.} {\bf B354}
(1995) 279.
\bibitem{pss2} G. Pradisi, A. Sagnotti and Ya.S. Stanev, {\sl Phys. Lett.} 
{\bf B356} (1995) 230.
\bibitem{wzw} E. Witten, {\sl Comm. Math. Phys.} {\bf 92} (1984) 455;\\
V.G. Knizhnik, A.B. Zamolodchikov, {\sl Nucl. Phys.} {\bf B247} (1984)
83.
\bibitem{fps} D. Fioravanti, G. Pradisi and A. Sagnotti, {\sl Phys. Lett.} 
{\bf B321} (1994) 349.
\bibitem{lew} D.C. Lewellen, {\sl Nucl. Phys.} {\bf B372} (1992) 654.
\bibitem{lud} A.W.W. Ludwig, {\sl Int. J. Mod. Phys.} 
{\bf B8} (1994) 347;\\
I. Affleck, {\sl cond-mat} 9512099, and references therein.
\bibitem{polc} J. Polchinski, {\sl Phys. Rev. Lett.} {\bf 75} (1995) 4724.
\bibitem{dbranes} E. Witten, {\sl hep-th} 9510135, 9511030;\\
P. Horava and E. Witten, {\sl hep-th} 9510209;\\
J. Polchinski and E. Witten, {\sl hep-th} 9510169;\\
C. Bachas, {\sl hep-th} 9511043;\\
J. Polchinski, S. Chauduri and C.V. Johnson, {\sl hep-th} 9602052,\\
and references therein.
\bibitem{ms} G. Moore and N. Seiberg, {\sl Phys. Lett.} {\bf B212}
(1988) 451,\\ {\sl Nucl. Phys.} {\bf B313} (1989) 16.
\bibitem{cardylew} J.L. Cardy and D.C. Lewellen, {\sl Phys.
Lett.} {\bf B259} (1991) 274. 
\bibitem{rst} Ya.S.Stanev, I.T.Todorov, L.K.Hadjiivanov, 
{\sl Phys.Lett.} {\bf B276} (1992) 87;\\
K.H. Rehren, Ya.S. Stanev and I.T. Todorov, 
{\sl Comm. Math. Phys.} {\bf 174} (1996) 605.
\bibitem{nsw} K.S. Narain, {\sl Phys. Lett.} {\bf B169} (1986) 41;\\
K.S. Narain, M.H. Sarmadi and E. Witten, {\sl Nucl. Phys.} 
{\bf B279} (1987) 369.
\bibitem{ciz} A. Cappelli, C. Itzykson and J.B. Zuber, {\sl Comm. Math. Phys.}
{\bf 113} (1987) 1.
\bibitem{pz} V. Petkova and J.B. Zuber, {\sl Nucl. Phys.} {\bf B438} 
(1995) 347.
\bibitem{ver} E. Verlinde, {\sl Nucl. Phys.} {\bf B300} (1988)
360.
\end{thebibliography}
\end{document}